\documentstyle[preprint,prb,eqsecnum,aps]{revtex}

\input mssymb
\begin{document}

\title{Collective excitation spectrum of a disordered Hubbard model}
\author{Yolande H. Szczech, Michael A. Tusch and David E. Logan}
\address{University of Oxford, Physical and Theoretical Chemistry Laboratory,
South Parks Road, Oxford OX1 3QZ (U. K.)}
\maketitle
\begin{abstract}
We study the collective excitation spectrum of a $d=3$ site-disordered
Anderson-Hubbard model at half-filling, via a random phase approximation (RPA)
about broken-symmetry, inhomogeneous unrestricted Hartree-Fock (UHF) ground
states. We focus in particular on the density and character of 
low-frequency collective excitations in the transverse spin channel. 
In the absence of disorder,
these are found to be spin wave-like for all but very weak interaction 
strengths, extending down to zero frequency and separated from a Stoner-like 
band, to which there is a gap. With disorder present, a prominent
spin wave-like band is found to persist over a wide region of the 
disorder-interaction phase plane in which the mean-field ground state is a
disordered antiferromagnet, despite the closure of the UHF single-particle 
gap. Site-resolution of the RPA excitations leads to a microscopic 
rationalization of the evolution of the spectrum with disorder and interaction
strength, and enables the 
observed localization properties to be interpreted
in terms of the fraction of strong local moments and their
site-differential distribution.
\end{abstract}

\section{Introduction}

Central to the understanding of any system with long-ranged order is a
description of its collective excitations. 
If the Hamiltonian has a continuous symmetry which is broken in the
ground state, their spectrum is gapless (as implied by the existence
of Goldstone modes) and they dominate low-temperature
thermodynamic properties.

One of the most familiar examples of such a system is
the Heisenberg model, whose collective
excitations are pure spin waves. With $S=\frac{1}{2}$, antiferromagnetic (AF)
nearest-neighbour interactions and on a bipartite lattice, the ground
state of the Hamiltonian has 
long-ranged magnetic order in
three dimensions $d$\cite{kls}. While no exact
solution exists for $d>1$, a successful and longstanding 
approach to the ground state
and collective excitations of the model is given by linear spin-wave
theory (LSW)\cite{lsw1}, in which particle-hole pairs created out of
the N\'eel mean-field 
ground state are treated as bosons. 
The theory
becomes exact in the limit $d\rightarrow\infty$, and $1/d$
corrections can be systematically incorporated\cite{lsw2,wang}; but even
in $d=2$, LSW estimates for the
sublattice magnetization, spin wave velocity and ground state energy
are in good agreement with quantum Monte Carlo calculations (QMC)
(see e.g. Ref. \onlinecite{lsw2} for a comprehensive review). 

It is well known that the AF Heisenberg model represents the strong
coupling limit of the half-filled Hubbard model\cite{concepts}. At
finite interaction strengths $U$, the mean-field ground state remains
the N\'eel state (albeit with self-consistently determined local
moment magnitudes) and gives a qualitatively sound description
of the true ground state in $d\geq 2$. The collective excitation
spectrum is however considerably more complex than that of the
Heisenberg model, since its low-frequency spin wave excitations
are mixed with incoherent, Stoner-like processes, leading to 
$O(N^2)$ particle-hole excitations in contrast to the $N$ spin waves
which alone survive in the strong-coupling limit.

The generalization of LSW to finite interaction strengths  is given by the random phase
approximation (RPA) about the fully unrestricted Hartree-Fock (UHF) saddle
point. Although  naively thought of as a weak-coupling theory, it
reduces precisely to LSW in the strong coupling limit, an observation
which has led a number of authors to conclude it to be a sound way of
probing the collective excitations of the Hubbard model over a wide
range of interaction strengths: for example, such an approach has
been successfully used to investigate the transverse spin excitation spectrum
of the Hubbard model both in $d=1$ \cite{bishop} and $d=2$ (see
e.g. Refs \onlinecite{singh,swz,sen}).

In the present work we consider initially the collective excitation
spectrum of the Hubbard model on a simple cubic lattice. 
This is shown to possess two distinct bands for all interaction
strengths $U$: a low-energy spin wave-like band corresponding primarily to
orientational fluctuations of the local moments, extending down to
zero frequency and containing two Goldstone modes; and an upper
band of Stoner-like excitations, to which there is a gap  
due to Fermi-surface nesting in the corresponding single-particle spectrum
\cite{penn}. The latter involve a significant degree of charge
transfer character (site double-occupancy) and are projected out in
the large $U$ limit, leaving solely the
low-frequency excitations, which become the pure spin waves of the
Heisenberg model. It is found that, despite mixing between the two
bands at finite $U$, the low-frequency excitations retain a strong
degree of spin wave character down to weak interaction strengths.

A problem about which very little is known, even at a qualitative
level, is the effect of disorder on the collective excitation 
spectrum. In a disordered
system, the collective excitations are governed by the interplay of a
number of complex processes, including strong correlations
of particle-hole pairs, Anderson localization of the
underlying single-particle states,
and the non-trivial statistics associated with the distributions of
local charges and magnetic moments arising from inhomogeneity in the site
environments.

Even at the mean field level of UHF, the combined effects of disorder
and electron interactions in leading to strongly inhomogeneous ground
states yield rich behaviour, as shown by a recent study of
the zero-temperature UHF phase diagram of a site-disordered Anderson-Hubbard
model\cite{prb} (AHM). Further, at RPA level, these ground states in
turn give rise to strongly inhomogeneous magnetic response
properties\cite{rowan,chipap}. In particular, disorder is found to
lead to a significant enhancement in local static susceptibilities
over those in the pure system, this enhancement being strongly
site-differential.

The behaviour of static and dynamic susceptibilities in both the spin and
charge channels ultimately reflects the character and distribution of
collective excitations about the ground state. The observed site-differential
enhancements mentioned above naturally raise questions both of how disorder
affects the excitation spectra of the model and of the spatial distribution
and localization characteristics of the 
excitations. 
In addition, the
stability of the long-ranged order present in the mean-field 
ground state towards
zero point spin fluctuations, and the low-temperature properties
of the model,are also governed by the RPA
excitations.

We here investigate of the RPA spectrum
of a $d=3$  Anderson-Hubbard model with Gaussian site-disorder, 
throughout the
disorder-interaction phase plane. 
In Section II, we discuss generic properties of the model and
describe the procedure for calculating and characterizing the
collective excitation spectrum at RPA level. After a brief discussion
of the pure Hubbard model in Section III, we focus in Section IV on
the AHM, concentrating on the distribution,
character and localization properties of the low-frequency  excitations, with particular
emphasis on the existence or otherwise of a spin wave-like band in the
transverse excitation spectrum at finite disorder. We conclude in
Section V with a discussion of the extent to which a
recently developed mapping of the excitations of the Hubbard model\cite{prbhub,prl}
onto those of an effective spin model, can be applied to the disordered
system.

\section{Collective excitation spectrum at RPA level}

The Hamiltonian we consider is given by 
\begin{equation}
H=\sum_{i,\sigma}\epsilon_i n_{i\sigma}-t\sum_{\langle ij\rangle,\sigma}
c_{i\sigma}^\dagger
c_{j\sigma}+{\scriptstyle\frac{1}{2}}
U\sum_{i,\sigma} n_{i\sigma}n_{i-\sigma}
\label{hubbard}
\end{equation}
where $t$ is the hopping matrix element, $U$ is the (repulsive) on-site Coulomb
interaction and 
the $\langle ij\rangle$ sum is  over nearest neighbour sites on a $d=3$ simple
cubic lattice. The site energies $\{\epsilon_i\}$ are drawn randomly from a common
Gaussian distribution $g(\epsilon)$ of variance $\Delta^2$
which, together with Eq. (\ref{hubbard}), specifies the Anderson-Hubbard model 
considered here; and we focus exclusively on 
half-filling. Making a spin-rotationally invariant UHF approximation to the
interaction term in Eq. (\ref{hubbard}) allows the Hamiltonian to be
expressed  as $H=H_0+H_1$ with
\begin{eqnarray}
H_0&=&\sum_{i,\sigma}\epsilon_i n_{i\sigma}-t
\sum_{\langle ij\rangle,\sigma}c_{i\sigma}^\dagger c_{j\sigma}\nonumber\\
& &+U\sum_i\{ {\scriptstyle\frac{1}{2}} 
\bar{n}_i n_i-2\bar{\bbox{S}}_i\cdot\bbox{S}_i\} ~~,
\label{uhfham}
\end{eqnarray}
\noindent where  $\bbox{S}_i$ is a spin-$\frac{1}{2}$ operator and the
overbar denotes an expectation over the UHF ground state. The
zero-temperature phase  diagram of $H_0$    
in the $(\Delta/t,U/t)$ plane has been discussed extensively in Ref. \onlinecite{prb};
 aspects of this study necessary for the present work will be summarized
briefly in \S IVA. Here we note that, while a rich variety of phases arises,
all mean-field magnetic ground states are found to be Ising-like
(i.e. fully collinear local moments, with
$\bar{S}_{ix}=0=\bar{S}_{iy}$ for all sites $i$), with $S_z^{\rm
tot}=\sum_i\bar{S}_{iz}=0$. 
Collective excitations about these ground states, obtained via
 the RPA, then decouple into two sets \cite{rowan}: for frequencies
$\omega\geq 0$ there exist $N^2/2$ spin excitations transverse to the local
moment $z$-axis, and $N^2/2$ longitudinal spin and charge excitations. In the
non-disordered ($\Delta/t=0$) Hubbard model there is a gap to
longitudinal spin
excitations for all $U/t$\cite{rowan}, reflecting the presence of a gap in
the single-particle density of states (DoS). By contrast the transverse
excitation spectrum is gapless (as reflected by the presence of Goldstone
modes). 
With disorder present, the dominant low-energy collective excitations
are again found in the transverse spin channel. We thus focus here on the transverse spin
spectrum, since it is these excitations which
govern measured susceptibilities (the existence of a spin-flop transition
implying that only transverse excitations are in practice probed \cite{chipap}), and
which are dominant in determining low-temperature thermodynamic properties.

\subsection{Determination of the transverse excitation spectrum}

The equations of motion for the RPA particle-hole excitations are 
 straightforwardly derived (see e.g. Ref. \onlinecite{fw}). However, solution of these
equations requires diagonalization of a $2N^2\times 2N^2$ non-Hermitian
matrix, a procedure  limited to impractically small system sizes.
An alternative ---and viable--- procedure is to exploit the fact that,
because the interaction term $H_1$ contains only on-site terms, expressions
for the RPA dynamic susceptibilities (derived either diagrammatically
or via linear response theory \cite{rowan}),  involve only matrices of
order $N$. Specifically, the  transverse spin susceptibility  

\begin{equation}
{\chi}^{-+}_{ij}(\omega)=i\int dt{\ } e^{i\omega t}\langle 0|{\cal T}\{S_i^-(t)S_j^+\}
|0\rangle_{\rm RPA} ~~~,
\label{chi}
\end{equation}

\noindent is given by 

\begin{equation}
\bbox{\chi}^{-+}(\omega)=\ ^0\bbox{\chi}^{-+}(\omega)\left[ {\bf 1}-U\,
^0\bbox{\chi}^{-+}(\omega) \right]^{-1} ~~,
\label{chieq}
\end{equation}
where the $N\times N$ matrix $^0\bbox{\chi}^{-+}(\omega)$ is the
corresponding UHF transverse susceptibility obtained 
by replacing $|0\rangle_{\rm RPA}$ by $|0\rangle_{\rm UHF}$ in Eq. (\ref{chi}). 
This is in turn given explicitly by
\begin{eqnarray}
^0\chi_{ij}^{-+}(\omega)&=&\sum_{\alpha>F>\beta}\left\{
{a_{i\alpha\uparrow}a_{j\alpha\uparrow}a_{i\beta\downarrow}a_{j\beta\downarrow} \over E_{\alpha\uparrow}-E_{\beta\downarrow}-
\omega-i\eta}\right.\nonumber\\
& &\left. +{a_{i\alpha\downarrow}a_{j\alpha\downarrow}a_{i\beta\uparrow}a_{j\beta\uparrow}\over
E_{\alpha\downarrow}-E_{\beta\uparrow}+\omega-i\eta}\right\}.
\label{uhfchi}
\end{eqnarray}
\noindent Here $F$ is the Fermi level, $\eta=0+$  and 
the $\{E_{\alpha\sigma}\}$ are the UHF single-particle energies, with eigenvectors
$|\Psi_{\alpha\sigma} \rangle=\sum_i a_{i\alpha\sigma}|\phi_{i\sigma}\rangle$ expanded in a site basis
(and which are pure spin-orbitals for Ising-like UHF
solutions \cite{prb}). We add in passing that the Fermi energy is pinned at 
$E_F=\frac{1}{2}U$ for all disorder strengths $\Delta\geq 0$, since the
familiar particle-hole symmetry characteristic of the non-disordered limit
is preserved for all $\Delta>0$ with the symmetric Gaussian $g(\epsilon)$\cite{prb}.

Eq. (\ref{uhfchi}) is just the Lehmann representation
for the  retarded polarization propagator appropriate to the UHF ground state,
with poles at the  Stoner excitation energies. When $\omega$ does not
coincide with a pole, this real symmetric matrix is diagonalized by an
orthogonal matrix ${\bf V}(\omega)$ with eigenvalues
$\{\lambda_\gamma(\omega)\}$. Most
importantly, as is evident from Eq. (\ref{chieq}), this transformation also
diagonalizes $\bbox{\chi}^{-+}(\omega)$ with 
corresponding eigenvalues
$\lambda_\gamma(\omega)/[1-U\lambda_\gamma(\omega)]$, viz:

\begin{equation}
\chi_{ij}^{-+}(\omega)=\sum_\gamma V_{i\gamma}{\lambda_\gamma(\omega)\over
1-U\lambda_\gamma(\omega)} V_{j\gamma}
\label{chivecs}
\end{equation}

\noindent RPA transverse collective excitations correspond to the
$\omega$-poles of $\bbox{\chi}^{-+}(\omega)$, which thus occur at
$1-U\lambda_\gamma(\omega)=0$, 
i.e. whenever an eigenvalue of $^0\bbox{\chi}^{-+}(\omega)$ crosses
$1/U$. Further, since $\chi_{ij}^{+-}(\omega)=\chi_{ji}^{-+}(-\omega)$,
a knowledge
of the (positive and negative) poles of $\bbox{\chi}^{-+}(\omega)$ alone is
sufficient to determine the full transverse spin excitation spectrum.

Two final points concerning the poles $\bbox{\chi}^{-+}(\omega)$ should be made
before proceeding. First, the broken symmetry of any magnetic UHF state
implies the presence of a zero-frequency Goldstone mode in both
$\bbox{\chi}^{-+}$ and $\bbox{\chi}^{+-}$, ensuring that the transverse
excitation spectrum is gapless. Second, the $\omega$-poles occur correctly on
the real axis only if the UHF ground state is fully stable against
particle-hole excitations (see e.g. Ref. \onlinecite{thouless}). 
An eigenstate of $H_0$ which is not a true minimum
on the UHF total energy surface leads to imaginary-$\omega$ excitations which
carry the system away from the saddle point, rendering it unstable. In
this regard, and even when the mean-field {\em solutions} are
Ising-like, use of a fully unrestricted and spin-rotationally
invariant Hartree-Fock approximation is an essential prerequisite,
since it ensures the mean-field states are minima.

In studying collective excitations at RPA level,
a pole search on $\bbox{\chi}^{-+}(\omega)$ ( Eqs (2.4,5))
 is powerful, since it allows 
study of much  larger system sizes  than direct solution of the RPA
equations. We now outline this procedure. 
From Eq. (\ref{uhfchi}), $^0\bbox{\chi}^{-+}(\omega)$ itself has poles at the UHF
excitation energies but, in the intervals between the poles, the
eigenvalues $\{\lambda_\gamma(\omega)\}$
of $^0\bbox{\chi}^{-+}(\omega)$ are continuous functions of $\omega$. 
For a given disorder realization, we thus determine the excitation spectrum as follows. 
(i) Solution of the UHF problem yields the UHF excitation energies and hence the
intervals over which the eigenvalues $\lambda_\gamma(\omega)$ are
finite. (ii)  Diagonalization of $^0\bbox{\chi}^{-+}(\omega)$ for $\omega$ at the
beginning and the end of each interval yields the number of eigenvalues which
have crossed $1/U$ ---and hence the number of RPA poles--- 
in each interval. (iii) Finer resolution may then be obtained  using,
for example, a bisection technique. 

\subsection{Character of the transverse excitations}

Important information concerning the character of a given excitation of
frequency $\omega_p$ can be obtained from the eigenvector coefficients
$\{V_{i\gamma}(\omega_p)\}$.

In the strong coupling limit $U/t\rightarrow\infty$, 
the resultant AF Heisenberg model contains $N$
transverse particle-hole excitations, corresponding to linear combinations of
solely on-site spin flips. These are just  the familiar pure spin
waves. At finite $U/t$, off-site spin flip  excitations also contribute to
the transverse spectrum, such that while there is spin-charge separation
(in the sense that the transverse spin and longitudinal/charge channels
decouple\cite{rowan}), the transverse excitations themselves 
possess some charge-transfer character.  It is particularly instructive to
evaluate the on-site spin flip (or spin wave) contribution to a given
excitation, as now considered. 
 
 The Lehmann representation for the diagonal elements
of  the RPA susceptibility is given generally by
\begin{equation}
\chi_{ii}^{-+}(\omega)=\sum_n{|\langle 0| S_i^-|n\rangle|^2\over {\cal
E}_{n0}-\omega-i\eta }+{|\langle 0|S_i^+|n\rangle|^2\over {\cal
E}_{n0}+\omega-i\eta} 
\label{lehmann}
\end{equation}
where $|0\rangle$ and $|n\rangle$ are RPA ground and excited states and  ${\cal E}_{n0}$
are the RPA excitation energies. From Eq. (\ref{chivecs}), 
${\chi}^{-+}_{ii}(\omega)$ is expressible in terms of the eigenvalues and
eigenvectors of $^0\bbox{\chi}^{-+}(\omega)$:
\begin{equation}
\chi_{ii}^{-+}(\omega)=\sum_\alpha V_{i\alpha}^2(\omega) 
{\lambda_\alpha(\omega)\over 1-U\lambda_\alpha(\omega)}
\label{chiii}
\end{equation}
To compare Eqs (\ref{lehmann}) and (\ref{chiii}), we analytically
continue Eq. (\ref{chiii})   by the replacement
$\lambda_\alpha(\omega)\rightarrow\lambda_\alpha(\omega+is)$,
where $s$ is an infinitesimal. Consider any particular pole,
$\omega_p$, of Eq. (\ref{chiii}), such that
$\lambda_\gamma(\omega_p)=1/U$. For frequencies
$\omega\simeq\omega_p$, $\chi_{ii}^{-+}(\omega)$ is dominated by the
$\alpha=\gamma$ term and is given to leading order by
\begin{equation}
\chi_{ii}^{-+}(\omega\simeq\omega_p)={V_{i\gamma}^2(\omega_p)\over
U^2\left(\partial\lambda_\gamma\over\partial\omega\right)_{\omega_p}(\omega_p-\omega-is)}
~~~.
\label{chiiiop}
\end{equation}
\noindent Comparison of Eqs (\ref{chiiiop}) and (\ref{lehmann}) implies
\begin{mathletters}
\begin{equation}
{\rm
sgn}\left(\partial\lambda_\gamma\over\partial\omega\right)_{\omega_p}
={\rm sgn}(\omega_p) ~~~,
\end{equation}
and
\begin{equation}
{\rm sgn}(s)={\rm sgn}(\omega_p)
\end{equation}
\end{mathletters}

\noindent (in order that the poles of $\bbox{\chi}^{-+}(\omega)$ occur
in the correct half-plane); together with the central result that, for
the given pole,

\begin{equation}
|\langle0|S_i^\rho|n\rangle|^2={1\over
U^2\left|\partial\lambda_\gamma\over\partial\omega\right|_{\omega_p}}
V_{i\gamma}^2(\omega_p)
\label{dlambda}
\end{equation}

\noindent where $\rho=-{\rm sgn}(\omega_p)$. Further, since the
eigenvector coefficients $\{V_{i\gamma}(\omega)\}$ are by construct
normalized, the sum rule

\begin{equation}
\sum_i|\langle0|S_i^\rho|n\rangle|^2={1\over
U^2\left|\partial\lambda_\gamma\over\partial\omega\right|_{\omega_p}}
\label{dlambda2}
\end{equation}

\noindent follows. That this sum is not unity simply
 reflects the fact that, at finite $U/t$, RPA transverse spin excitations have
finite weight outside the on-site spin flip subspace, i.e. matrix elements of
the form $\langle 0|c_{i\downarrow}^\dagger c_{j\uparrow}|n\rangle$ with $i\neq j$ are
non-zero; for example  it may be
shown that the general RPA normalization expressed in a site basis
implies that
$\sum_{ij} |\langle 0|c_{i\downarrow}^\dagger c_{j\uparrow}|n\rangle|^2 =1 $
for RPA excitations corresponding to positive-$\omega$ poles. Only in
the strong coupling limit $U/t\rightarrow \infty$ are these 
off-site excitations effectively projected out of the transverse spin spectrum.

The interpretation of the eigenvectors $\{{\bf V}_\gamma(\omega)\}$ is
therefore clear: from Eqs (\ref{dlambda}) and (\ref{dlambda2}) the
set of coefficients $\{V_{i\gamma}(\omega_p)\}$ corresponding to a particular
$\lambda_\gamma(\omega_p)=\frac{1}{U}$ describe the spatial distribution of the on-site
spin flip component of the collective excitation, while the prefactor
$\Omega(\omega_p)= 1/(U^2
\left|{\partial \lambda_\gamma/\partial\omega}\right|_{\omega_p})$ gives the weight of the RPA
excitation in the subspace of on-site spin flips.  Excitations with
$\Omega(\omega_p)\simeq1$ are predominantly spin waves,
while those with $\Omega(\omega_p)\ll 1$
have significant off-site `charge transfer' character. (Note that, to
generalize to a disordered
system, we use the term `spin wave' to denote any excitation with
$\Omega\simeq1$, regardless of its  $\bbox{q}$-resolution.)

The above interpretation may also be exploited  to define an inverse participation
ratio (IPR) for the collective excitations, in analogy to the familiar IPR
employed in the study of localization of single-particle states \cite{bell}, via
\begin{equation}
L(\omega_p)=\sum_i V_{i\gamma}^4(\omega_p)~~~ .
\label{ipr}
\end{equation}
\noindent Using Eq. (2.12), this may be expressed as
\begin{equation}
{L}(\omega_p) = {\sum_i |\langle 0|S_i^\rho|n\rangle|^4\over
\left(\sum_i|\langle 0|S_i^\rho|n\rangle|^2\right)^2} ~~~~~~~~: \rho=-{\rm sgn}(\omega_p)~~~.
\end{equation}
${L}(\omega_p)$ is on the order of the inverse of the number of sites
overlapped by the excitation at $\omega_p$: for an excitation receiving uniform 
contributions from  $m$ sites, $L(\omega)\sim 1/m$. In the thermodynamic limit,
$L(\omega)$ is thus zero for a delocalized excitation and non-zero for a
localized excitation. For a finite-sized system $L>0$ necessarily, requiring
the estimation of a threshold  IPR, appropriate to a given system size.
While for the single-particle excitations of the 
(non-interacting) Anderson model, such a threshold
IPR can be  addressed via finite-size scaling \cite{skinner}, we have no
information on the $N$-scaling of $L(\omega)$ in the present system. 
In \S IV, however, we argue that a qualitative understanding of the localization characteristics 
of transverse spin excitations may nonetheless be obtained.

\section{Excitation spectrum of the pure Hubbard model}

Before discussing the effects of disorder on the transverse excitation
spectrum of the Anderson-Hubbard model, 
we first consider briefly the important non-disordered limit. In this
case, for all $U/t>0$, the UHF single-particle spectrum
$D(E)=\frac{1}{2}\sum_{\alpha\sigma}\delta(E-E_{\alpha\sigma})$ has a band gap of magnitude
$U|\mu|$, with $|\mu|$ the self-consistent UHF local moment magnitude;
and in which lies the Fermi level $E_F=\frac{1}{2}U$\cite{penn}.
Consequently, the pure UHF particle-hole excitation spectrum consists
of a single Stoner
band beginning at $\omega=U|\mu|$,  with a maximum at $\omega\sim U$.
The full RPA spectrum of collective transverse spin excitations
consists, by contrast, of two distinct components (we consider
$\omega\ge 0$ throughout, since the spectrum is symmetric in
$\omega$). First, a low frequency band containing precisely $N$
excitations of predominantly spin wave character (discussed below),
and extending down to $\omega=0$ as implied by the presence of the
Goldstone modes for all $U/t>0$. Second, a high frequency band
containing $(N/2)(N-2)$ Stoner-like excitations, which closely
resembles the corresponding pure UHF transverse spin spectrum: it
begins at $\omega=U|\mu|$ ---the gap in the UHF single-particle
spectrum--- and has maximum spectral density for $\omega\sim U$. This
Stoner-like band is effectively eliminated as
$U\rightarrow\infty$. For all $U/t>0$, however, it may be shown to be
separated from the top of the low frequency band by a spectral gap,
i.e. there is a persistent separation of scales between the low energy
spin wave-like band and higher energy Stoner-like excitations. For
this reason we focus exclusively on the low-$\omega$ component of the
spectrum.

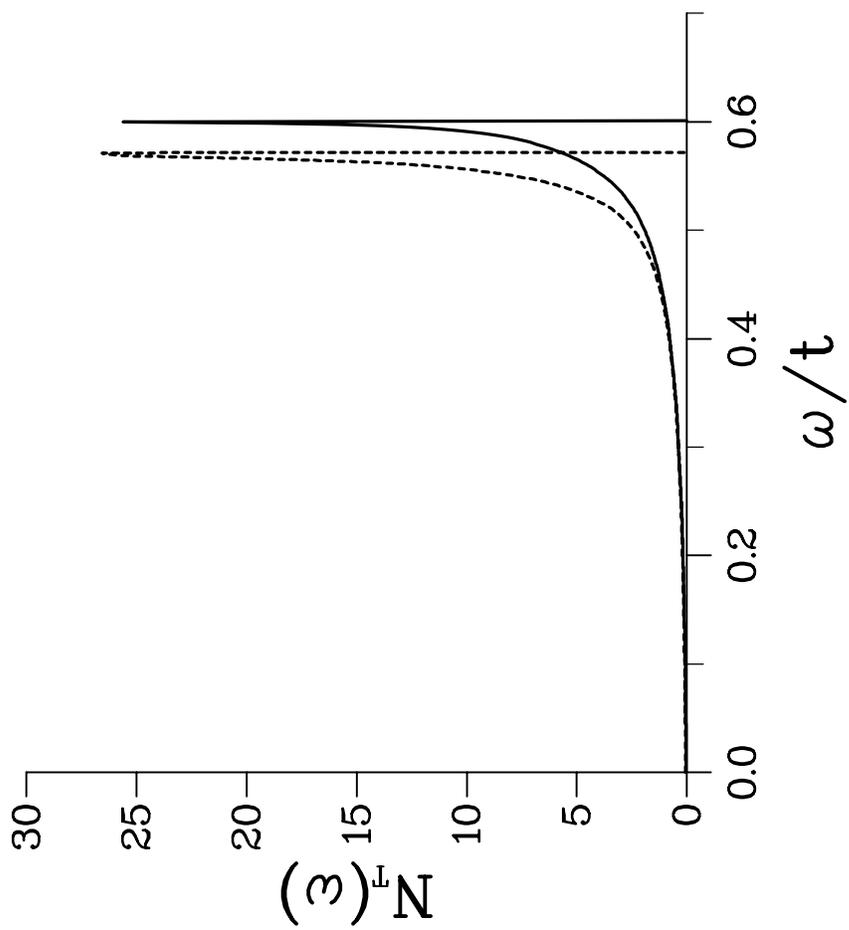
\begin{figure}
\caption{
Density of low-frequency RPA excitations $N_T(\omega)$ 
for the pure Hubbard model
with $U/t=20$ (solid line), compared to the linear spin wave spectrum of the
pure AF Heisenberg model with nearest-neighbour $J=4t^2/U$ (dashed line). 
}
\label{f7}
\end{figure}

The prescription of \S II, reformulated in a two-sublattice
basis\cite{singh} appropriate to the half-filled, non-disordered
model, allows the RPA spectrum to be determined; in practice we find
that a lattice of $32^3$ ${\bf q}$-points (corresponding to $N=32768$)
is required to give a sufficiently dense spectrum. Fig. \ref{f7}  shows the
resultant density of the lowest $N$ transverse spin excitations, given by
\begin{equation}
N_T(\omega)=N^{-1}\sum_{\omega_p\leq \omega_{\rm max}}\delta(\omega-\omega_p)
\end{equation}
where $\omega_{\rm max}$ is the frequency of the $N^{\rm th}$ excitation, for
large coupling $U/t=20$. Also shown is the corresponding spectrum for the AF
Heisenberg model with solely nearest-neighbour (NN) $J=4t^2/U$, given by $N_T^{\rm
LSW}(\omega)=N^{-1}\sum_{\bf q} \delta(\omega-\omega_{\bf q})$ with \cite{lsw1}
\begin{equation}
\omega_{\bf q}=\pm 3J\sqrt{1-\gamma_{\bf q}^2}
\end{equation}
where $\gamma_{\bf q}={1\over3}\sum_{\alpha=1}^3 \cos(q_\alpha a)$ and
$a$ is the lattice constant. 
The spectrum of low-frequency transverse  spin excitations of the Hubbard model thus
closely resembles that of the NN Heisenberg model at large $U/t$ ---as
expected, since as $U/t\rightarrow\infty$ the former maps exactly onto
the latter--- although as seen there is a not inappreciable
quantitative difference even for $U/t=20$.

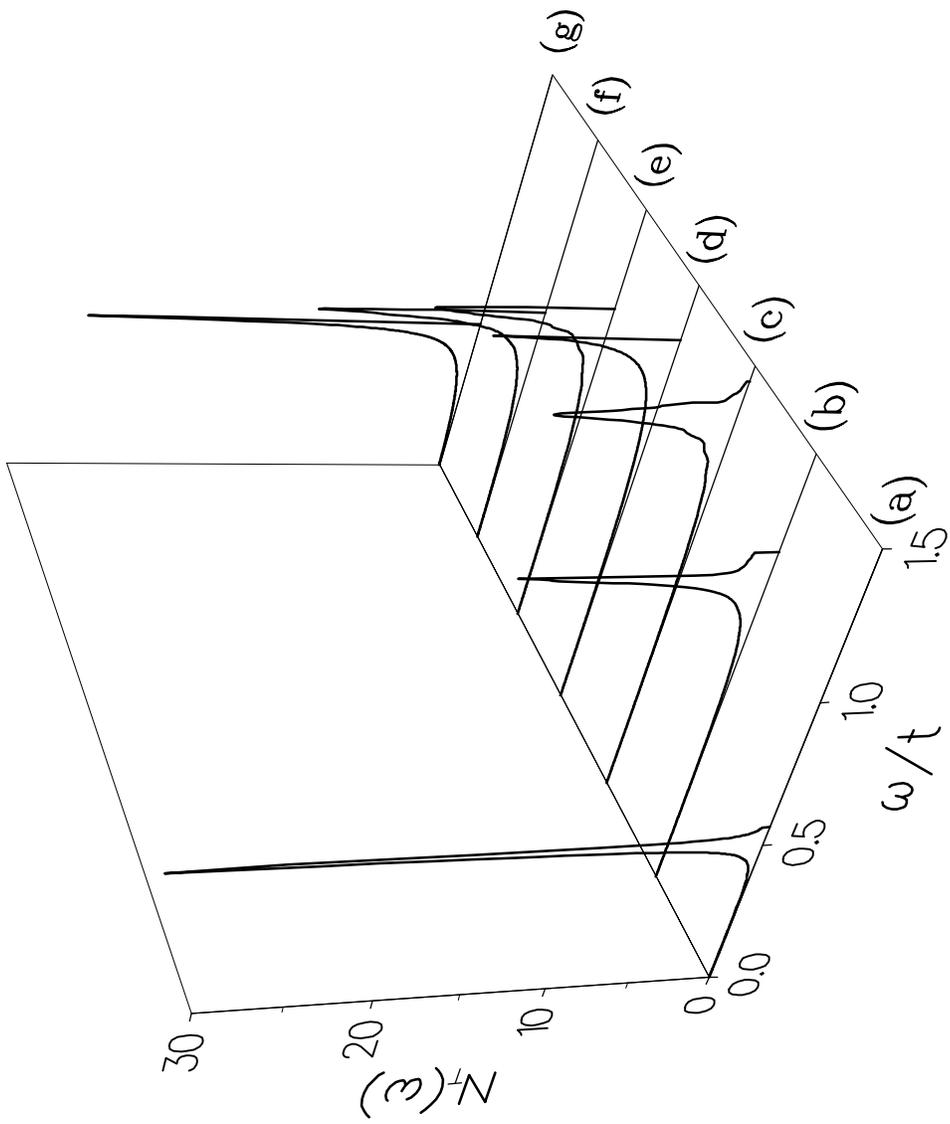
\begin{figure}
\caption{
Density of the $N$ lowest-frequency
RPA excitations $N_T(\omega)$ for the pure Hubbard
model for  increasing  interaction strengths: $U/t=2$ (a), 3 (b), 4 (c), 6
(d), 8 (e), 12 (f), 20 (g).
}
\label{f8}
\end{figure}

Less predictably, this qualitative similarity persists down to weak
coupling. However, as $U/t$ is decreased, the energy scale for low-frequency RPA excitations
rapidly deviates from that of the  NN Heisenberg model with $J=4t^2/U$.
Fig. \ref{f8} shows the effects of gradually decreasing $U/t$ down to
$U/t=2$. Initially, the spin-wave band is pushed to {\it higher} energy, 
consistent with the strengthening of NN exchange couplings  and
reflecting also  the increasing importance of beyond-nearest neighbour 
couplings as shown in a recent paper \cite{prbhub}. Below $U/t\sim
5$, however, this band starts to migrate to {\em lower} energies, 
reflecting an increasing repulsion with the approaching Stoner-like band. 

To investigate quantitatively the spin-wave character of the
excitations, we calculate
$\Omega(\omega_p)=1/(U^2|\partial \lambda/\partial \omega|_{\rm \omega_p})$
for $\omega_p\leq \omega_{\rm max}$. The lowest excitation is of
course the Goldstone mode corresponding to a global spin rotation, and
is a pure spin wave for all $U/t>0$: $\Omega(\omega_p=0)=1$. Since
excitations at the upper edge of the spin wave-like band are naturally
most strongly influenced by the encroaching Stoner-like band, it is
found that $\Omega(\omega_p)>\Omega(\omega_{\rm max})$ for all
$\omega_p<\omega_{\rm max}$. The first column of Table I thus shows the
$U/t$-dependence of $\Omega(\omega_{\rm max})$. From this it is seen that
for $U/t\gtrsim 3$ the excitations are predominantly spin waves -- in
fact for $U/t\gtrsim 5$ they are in essence pure spin waves,
$\Omega=1$ ---commensurate with a significant gap between the spin
wave and Stoner-like bands. For $U/t\lesssim 2-3$, this gap diminishes
rapidly and there is a substantial loss of spin wave character for
excitations with $\omega\simeq\omega_{max}$. Nonetheless, as is evident
from the above, the low energy transverse spin excitations of the
Hubbard model remain predominantly spin waves for all but the lowest
$U/t$.

\section{Anderson-Hubbard model}

Before considering  the effects of disorder on the RPA
spectrum, we first briefly outline relevant  properties of the inhomogeneous UHF
ground state,  since it is particle-hole excitations out of this state which
form the basis for the RPA collective modes; a full account of the
mean-field phase diagram is given in Ref. \onlinecite{prb}.

\subsection{Mean-field ground state}

On the introduction of disorder, UHF solutions remain Ising-like, with $S_z^{\rm
tot}=0$. Three distinct magnetic phases
are found: disordered paramagnetic (P), antiferromagnetic (AF) and
spin glass-like (SG) phases. Both metallic (M)  and insulating (I) phases also
occur, with the dominant metal-insulator transition (MIT)
M$\rightarrow$gapless I, driven by Anderson localization of single-particle
states at the Fermi energy $E_F=\frac{1}{2}U$. We now consider two specific paths
across the disorder-interaction phase 
plane, which will be employed in the next section to illustrate the combined
effects of $\Delta$ and $U$ on the RPA transverse spin  spectrum. 

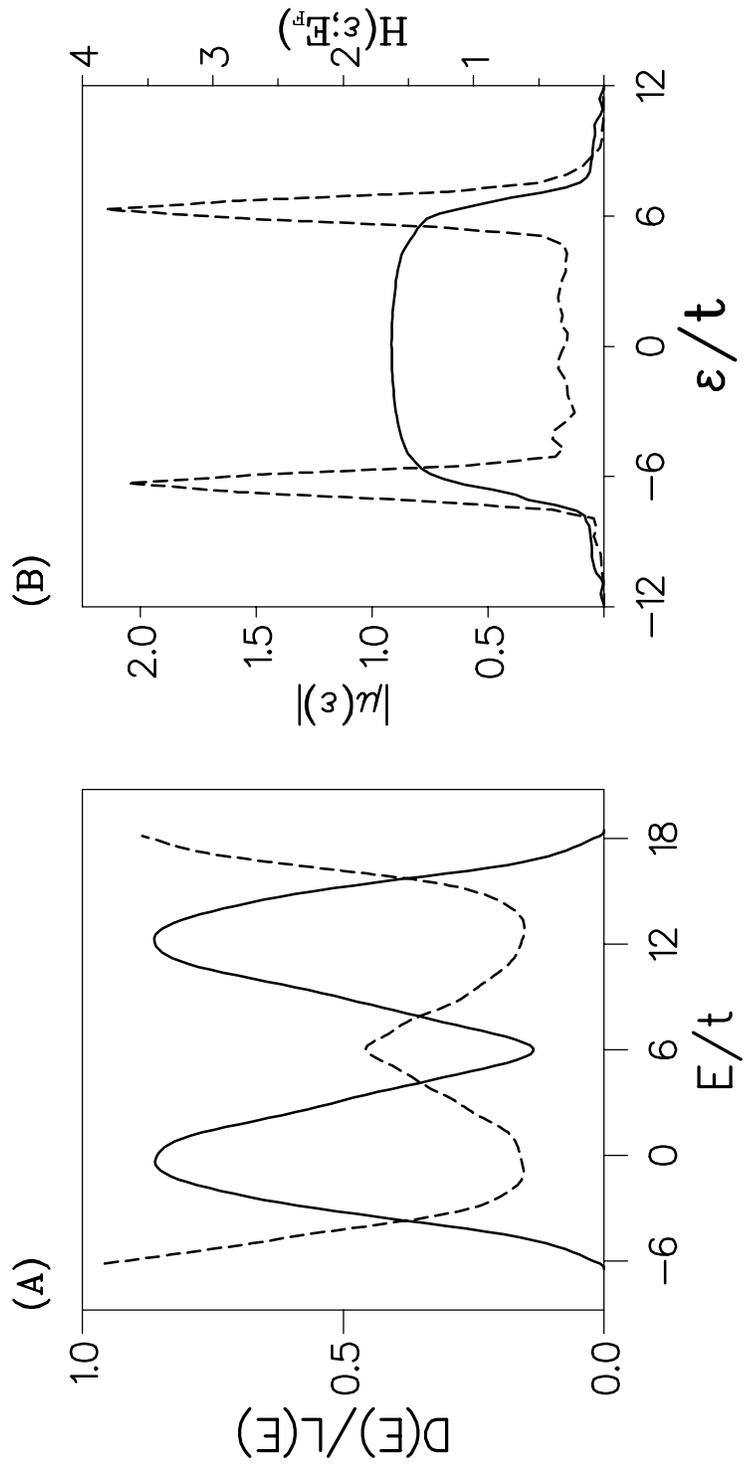
\begin{figure}
\caption{
For $U/t=12$ and $\Delta/t=3$, (A) Disorder-averaged
density of UHF single-particle states D(E) vs $E/t$ (solid line), where  the
unperturbed ($U=0$) bandwidth is $12t$; IPR $L'(E)$ for $N=512$
site systems (dashed line);
(B) Corresponding site-resolved local moment distribution $|\mu(\epsilon)|$ 
vs site energy $\epsilon/t$ (solid line,
left-hand scale), $H(\epsilon;E_F)$ as defined in text 
(dashed line, right-hand scale).
}
\label{f5}
\end{figure}

First, consider
fixing $U/t=12$ and varying disorder, $\Delta/t$. As $\Delta/t$ is increased
from zero, disorder leads to broadening and eventual overlap of the
Hubbard bands in the UHF single-particle spectrum, such that $E_F$
lies within a pseudogap in the total DoS 
$D(E)$. Studies of the IPR $L'(E)= \sum_i|a_{i\alpha\sigma}|^4$ 
for single-particle states of energy $E_{\alpha\sigma}=E$, with $E \sim
E_F$  \cite{prb}, imply that the system passes directly from a gapped to gapless
insulator in this region of the $(\Delta/t,U/t)$ phase plane (at
$\Delta/t\simeq 2.5$). Fig \ref{f5}(A)
shows $D(E)$ and the IPR profile $L'(E)$ averaged over $N=512$ site disorder
realizations  for $(\Delta/t,U/t)=(3,12)$.  A pronounced pseudogap is
evident in $D(E)$, in which $E_F$ lies. Correspondingly, $L'(E)$ shows a
maximum at $E=E_F$ and minima for states in the centre of the Hubbard
bands where $D(E)$ is maximal. Using a threshold IPR 
appropriate to the chosen system size\cite{prb} shows that Fermi level
states are localized: the system is a gapless Mott-Anderson insulator
(and remains thus with increasing $\Delta/t$ at the chosen $U/t=12$).
Analysis of the Fourier transform of the local moment
$S_z(\bbox{k})=N^{-1}\sum_i \mu_i{\rm e}^{i{\bf k}\cdot{\bf R}_i}$
shows however that the
system remains antiferromagnetically ordered: although increasingly `dirtied' by disorder,
$|S_z(\bbox{\pi})|\simeq |\mu|$ (where the 
mean moment per site $|\mu|=N^{-1}\sum_i|\mu_i|$), with
little weight at other $\bbox{k}$-vectors. Note however that while the phases of
the moments remain locked in AF alignment, the distribution of local moment
{\em magnitudes} over the sites is highly disordered. This is 
evident from the site-resolved distribution of local moment
magnitudes, $|\mu(\epsilon)|=N_\epsilon^{-1}\sum_{i:\epsilon_i=\epsilon}|\mu_i|$, with
$N_\epsilon$ the number of sites with $\epsilon_i=\epsilon$; as shown in Fig \ref{f5}(B), again for
$(\Delta/t,U/t)=(3,12)$. Sites with energies $|\epsilon_i|\lesssim \frac{1}{2}U$,
disposed randomly on the lattice, carry strong atomic-like moments, while
sites with lower or higher site energies have $|\mu_i|\simeq 0$, being
respectively largely doubly occupied by electrons or empty. This
profile is found to vary
little with disorder for fixed $U/t=12$, aside from minor erosion of local
moment magnitudes:  the primary effect of $\Delta/t$ is  to
determine, via the site-energy distribution $g(\epsilon)$, the fraction of sites
(around $\epsilon=0$) which carry strong local moments. 

Also shown in Fig. \ref{f5}(B) is a measure of the distribution of 
quasiparticle states of {\it given} energy $E$ over the sites:
$H(\epsilon;E)$, given by\cite{prb} 
\begin{mathletters}
\label{heps}
\begin{equation}
H(\epsilon;E)={g(\epsilon)D(\epsilon;E)\over D(E)} ~~~,
\end{equation}
where
\begin{equation}
D(\epsilon;E)={\scriptstyle\frac{1}{2}} N_\epsilon^{-1}\sum_{i:\epsilon_i=\epsilon}
\sum_{\alpha\sigma}|a_{i\alpha\sigma}|^2\delta(E-E_{\alpha\sigma})
\end{equation}
\end{mathletters}
is the partial density of states for sites of bare site energy $\epsilon$.
$H(\epsilon;E_F)$ is thus the quantum probability density that electrons in
Fermi-level quasiparticle states will be found on sites of bare site energy
$\epsilon$. As discussed in Ref. \onlinecite{prb}, and exemplified in
Fig. \ref{f5}(B), $H(\epsilon;E_F)$ 
is peaked strongly at the boundaries of the local moment distribution
$|\epsilon|\simeq \frac{1}{2}U$; i.e.  sites with strong local moments
participate but weakly in Fermi-level quasiparticle states, which are in
contrast dominated by sites with bare site energies close to the local moment
boundaries. Sites with strong local moments, around $\epsilon=0$, are in turn found
to contribute mainly to quasiparticle states with $E\simeq 0$ and $E\simeq
U$, i.e. around the maxima in the single-particle spectrum.

The second path considered is
fixing the disorder at $\Delta/t=3$, and decreasing the scaled
interaction strength $U/t$ from $U/t=12$. Note that, since the
$U/t\rightarrow\infty$ limit of the disordered Hubbard model for any $\Delta/t$
is the {\em non-disordered} AF Heisenberg model, decreasing $U/t$ may be
thought of as effectively increasing (by `switching on') the role of disorder.
As $U/t$ is reduced, the fraction of sites at the local moment boundaries
$\sim g(\pm\frac{1}{2}U$) ---which give the dominant contribution to
Fermi-level states (Fig. \ref{f5}(B))--- increases. The system is ultimately driven metallic at
a critical $U/t\simeq 7.2$\cite{prb}. However, magnetic ordering remains characterized by a
strong peak in $S_z(\bbox{k})$ at $\bbox{k}=\bbox{\pi}$ until
$U/t\simeq 2.5$, whereupon a sharp\cite{rowan}
transition occurs to a spin glass-like state, with $S_z(\bbox{k})$ receiving
essentially uniform $O(1/N)$ contributions from many $\bbox{k}$
vectors; see Ref. \onlinecite{rowan}. In this phase, although the mean
local moment per site $|\mu|$ is typically small ($\sim 0.05$),
relatively strong local moments are in fact found on a small fraction of
sites with $|\epsilon|\lesssim\frac{1}{2}U$ for a given disorder
realization. As discussed in Ref. \onlinecite{prb}, this reflects the 
disorder-induced production of statistically rare local environments in
which atomic-like moments are not significantly eroded by electron
hopping processes.

\subsection{Collective excitation spectrum of the Anderson-Hubbard model}

We now turn to the effects of disorder on the transverse RPA excitation
spectrum. The principal questions we seek to address are:
\begin{itemize}
\item What is the effect of  disorder on the density of transverse spin
excitations? Does a low-energy spin wave-like band exist for finite disorder,
and what is the character of the associated excitations?

\item What is the effect of disorder on the localization characteristics of
low-lying collective excitations?

\item What is the distribution of transverse spin  excitations of given energy over the sites,
and how does this relate to the inhomogeneous distribution of local moments
in the mean-field ground state?
\end{itemize}

With disorder present, there are no simplifying symmetries which allow
very large system sizes to be studied, as in the $\Delta=0$ case. However, this
lack of symmetry ---specifically, the absence of degeneracies in the UHF
spectrum--- enables reasonable statistics to be gained by sampling many
disorder realizations for  system sizes much smaller than those required for
$\Delta=0$. Numerical evaluation of the 
entire transverse excitation spectrum ---including the Stoner-like
excitations ---remains nonetheless  a large problem: a full
pole search for a system of $N$ sites requires diagonalization of a $N\times
N$ matrix in the vicinity of $\frac{1}{2}N^2$ poles, and is therefore
$O(N^4)$ for each disorder realization; an ensemble of which needs
subsequently to be sampled. For calculations of the full transverse
spin spectrum we have thus chosen $N=64$ site systems, which are found
to be sufficient for reproduction of its general features; while for
more detailed study of the low energy portion $N=216$ is employed. 

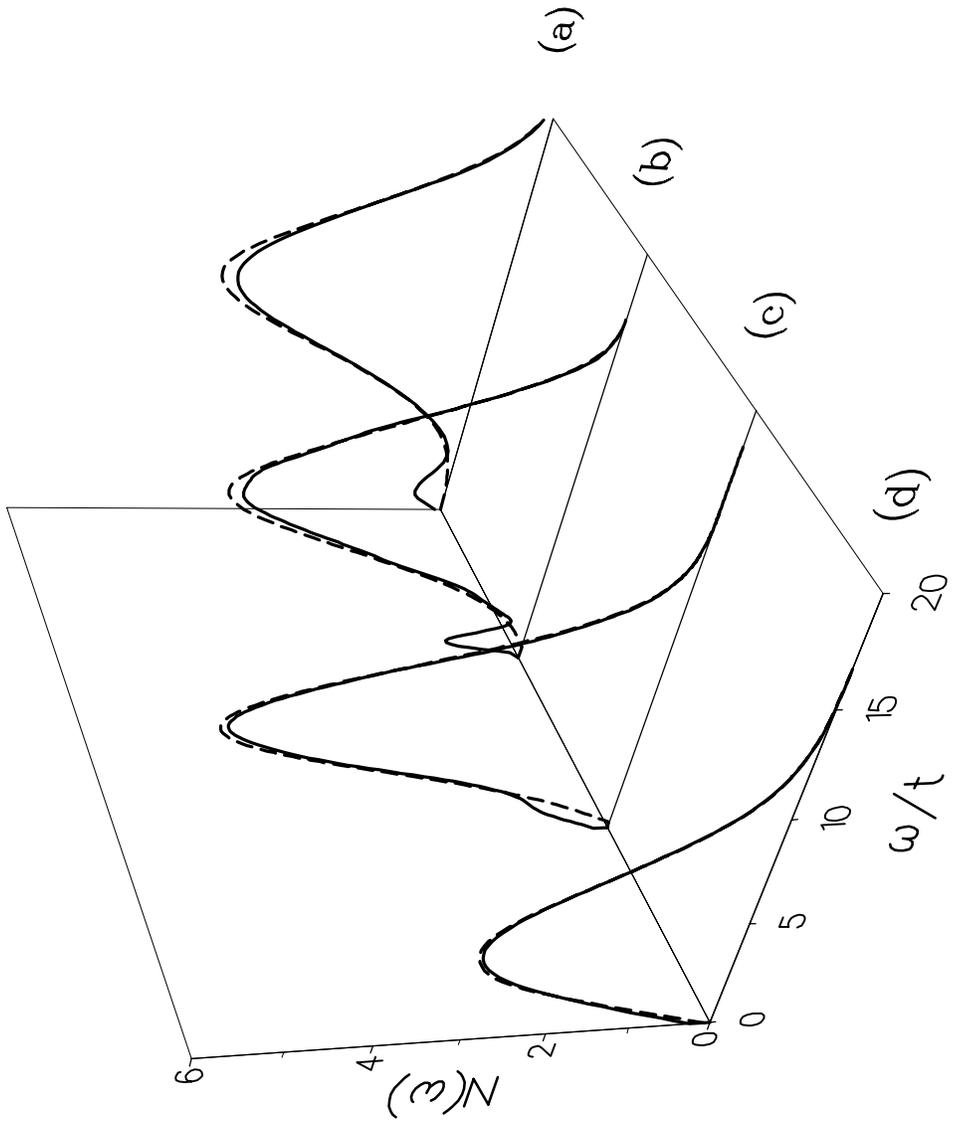
\begin{figure}
\caption{
Full  transverse excitation spectrum for the Anderson-Hubbard model 
at RPA level (solid line) for $N=64$ averaged over many disorder realizations, 
compared to the corresponding UHF Stoner spectrum (dashed) for
fixed disorder $\Delta/t=3$ and $U/t=12$ (a), 9 (b), 6 (c), 3 (d)
}
\label{f9}
\end{figure}

For fixed disorder $\Delta/t=3$ and various values of the interaction
strength $U/t\leq12$, Fig. \ref{f9} shows the full RPA density of transverse
spin excitations, together with the corresponding UHF particle-hole
excitation spectrum (the `pure Stoner' spectrum).  At $U/t=12$, where
the mean-field  ground state is
a gapless  AF insulator, a pronounced spin wave-like band is again evident in the RPA
spectrum at low $\omega$, as for the non-disordered limit  (although
the pure Stoner spectrum  now
extends down to $\omega=0$);
 the remainder of the RPA spectrum closely resembles the Stoner
spectrum. On decreasing
$U/t$, the pure Stoner  spectrum shifts to lower energy, 
since  the separation between the maxima in the corresponding single-particle
spectrum  $D(E)$ decreases and 
 the pseudogap at $E_F$  is gradually eliminated. As a result,  the spin wave-like band
is gradually shifted to lower energy (`softened') by, and ultimately
absorbed into, the Stoner-like band. By $U/t=3$, close
to the AF--SG border, the pure Stoner and RPA spectra appear virtually
indistinguishable.

On fixing $U/t$ and increasing disorder, a somewhat different  picture
emerges. Fig. \ref{f10}   shows the RPA and pure Stoner spectra for $U/t=12$ and
various values of $\Delta/t$. At low disorder $\Delta/t=1$, for which 
the UHF ground state is
a AF Mott-Hubbard (gapped) insulator,  a spin wave-like band, well separated from the
remainder of the excitations, is clearly distinguishable. Closer
resolution shows that 
this band is in fact shifted to  slightly {\em higher} frequency 
compared to the non-disordered case, despite the intuitive expectation that
disorder should lead to a `softening' of the spin waves.  This upward  shift is
correspondingly  manifest in
an initial decrease in the uniform static susceptibility  
$\chi_u=N^{-1}\sum_{ij}\chi^{-+}_{ij}(0)$ upon the introduction of disorder:
as follows from the Lehmann representation of the susceptibility
Eq. (\ref{lehmann}), a decrease in density of low-frequency excitations
with significant weight in the on-site spin flip subspace leads  to a
decrease in $\chi_u$. 
On increasing disorder further, the UHF single-particle gap closes and
the pure Stoner spectrum acquires weight down to $\omega=0$, Fig. \ref{f10}. 
The maximum in the Stoner-like band in the RPA spectrum
---occurring at $\omega\sim U$--- changes little with disorder, but
its disorder-induced broadening
 increasingly repels the spin wave-like band which is thus by contrast
shifted to {\it lower}
frequency (see Fig. \ref{f10}).
This is the `softening' of spin waves with disorder, which  in turn is
reflected in an increase in the uniform static susceptibility $\chi_u$.
 Fig. \ref{fchi} shows the disorder-averaged $\chi_u$ vs $\Delta/t$
for $U/t=12$ (and we add that in the non-disordered limit, the
resultant $\chi_u$ for the $N=64$ site system reproduces accurately
the $N\rightarrow\infty$ value, which can be obtained
analytically). The above behaviour is clearly demonstrated: 
after an initial slight decrease, $\chi_u$ progressively
increases with increasing disorder as the spin wave-like  band moves to lower frequencies.
Note further that increasing disorder for fixed $U/t$ has a considerably less
pronounced effect on the spin wave-like band than decreasing $U/t$ for fixed
disorder: decreasing interaction strength shifts the maximum in the
Stoner-like band to lower energy (Fig. \ref{f9}), while increasing $\Delta/t$ 
by contrast merely broadens the band, as shown in Fig. \ref{f10}.

The effect of disorder on the position of the spin wave-like band also
provides a natural explanation for the strongly site-differential
enhancements of local susceptibilities with disorder, as 
found in a recent RPA study of the present model\cite{chipap}. Specifically,  the total
unform susceptibility is deconvoluted via
\begin{equation}
\chi_u=\int d\epsilon g(\epsilon)\chi(\epsilon)
\end{equation}
where $\chi(\epsilon)=N_\epsilon^{-1}\sum_{i:\epsilon_i=\epsilon}\chi_i$ with
$\chi_i=\sum_j\chi_{ij}(0)$. $\chi(\epsilon)$ is thus the mean local susceptibility
for sites of {\it given} site energy $\epsilon$. In
ref. \onlinecite{chipap} (Fig. 2) it was found that an increase in disorder 
leads to a strong enhancement of $\chi(\epsilon)$ for sites in the local moment
range $|\epsilon|\lesssim\frac{1}{2}U$, while sites outside this
range, whose susceptibilities have typical Pauli-like values, are by
contrast only weakly affected by disorder. A site-differential
analysis of the RPA excitations forming  the spin wave-like band, in  analogy to
$H(\epsilon;E)$ for single-particle states (Eq. (\ref{heps})), shows that these
excitations have weight almost exclusively on the strong local 
moment-carrying sites, and receive little contribution from sites outside the local
moment range (consistent with their interpretation as `spin wave'-like). As
discussed above, the principal effect of increasing disorder is ultimately to
shift the spin wave-like band to low frequency, while the Stoner-like band is
merely broadened.  The disorder-induced `softening' of the spin wave-like
band therefore manifests itself almost exclusively in 
$\chi(\epsilon)$ for strong moment-carrying sites, as found in Ref. \onlinecite{chipap};
and has little effect on 
susceptibilities outside the local moment range, which are instead
dominated by Stoner-like excitations.

\subsection{Character of the transverse excitations}

From the preceding section, it is clear (Figs \ref{f9},\ref{f10})  that a low-energy feature in the
density of  RPA transverse spin excitations indeed exists in the AF region of the
$(\Delta/t,U/t)$ phase plane. However,  the preceding analysis by
itself gives no
information on the spin wave-like {\em character} of the excitations which
form this band, or  with which sites they are associated. To probe
the nature of these low-energy excitations, we have performed pole searches
using $N=216$ sites for many disorder realizations at a selection of points
in the $(\Delta/t,U/t)$ plane, focusing (as in the non-disordered
limit) on the lowest $N$ excitations, with
$\Omega(\omega_p)=1/(U^2|\partial
\lambda/\partial\omega|_{\omega=\omega_p})$  
evaluated at each RPA pole. For illustration, the second and third columns
respectively of Table I show
$\Omega(\omega)$ averaged over the lowest $N$  RPA excitations  for (a) fixed
$\Delta/t=3$ and various interaction strengths  $U/t$; 
and (b) for fixed $U/t=12$ as a function of disorder.  

For fixed $U/t=12$, it is seen that while increasing disorder leads to a
diminution of spin wave character, the effect is relatively weak and the
low energy transverse spin excitations retain strong spin wave
character. This is consistent with the fact that the system remains an
AF insulator, and with the persistence (Fig. 5) of the pronounced low
energy spin wave-like band in the transverse spin spectrum. By contrast,
reducing $U/t$ at fixed disorder leads to a more pronounced reduction 
in spin wave character, consistent with the progressive erosion of the
spin wave-like component in the RPA transverse spin spectrum (Fig. 4) 
discussed above. Again, this is physically natural: with decreasing 
interaction strength the system undergoes a (gapless) 
insulator$\rightarrow$metal transition at $U/t\simeq 7.2$ and an
increase in off-site character in the low-energy spin excitations is
thus to be expected. Note however that, even at low $U/t$, the RPA 
excitations have substantially more spin 
wave character than pure Stoner excitations 
in the same frequency range, as may be
verified by considering the weight of the latter in the on-site spin flip
subspace, given by
\begin{equation}
\Omega_{\rm HF}(\omega_p^0)=
\sum_{i\sigma}\sum_{\alpha>\beta}
a_{i\alpha\sigma}^2 a_{i\beta-\sigma}^2\delta(\omega_p^0-[E_{\alpha\sigma}-E_{\beta-\sigma}])
\end{equation}
This is found to be an order of magnitude smaller than the
corresponding $\Omega(\omega)$ for low-frequency RPA excitations.

\subsection{Localization characteristics of low-$\omega$ excitations}

Loss of translational symmetry on the introduction of disorder is expected to
lead to localization of some or all collective excitations, 
in analogy to Anderson localization
of single-particle states. 
Consider the IPR $L(\omega_p)$ of a collective transverse spin excitation 
of energy
$\omega_p$, as defined in Eq. (\ref{ipr}). To distinguish
between localized and delocalized excitations 
a threshold IPR appropriate to a given system size should 
be established, such that excitations below (above) this
value are deemed extended (localized). 
In principle, this could be obtained by finite-size scaling, but in
practice we are limited to system sizes on the order of $N=216$. For
single particle states of the non-interacting Anderson model,
finite-size scaling yields a threshold IPR $L_c'\sim 0.1$ for $N=216$
sites. We correspondingly adopt a threshold $L_c\sim 0.1$ for collective
excitations of the Anderson-Hubbard model. This value itself should not
of course be taken very seriously; it simply serves as a useful
qualitative guide in the following discussion.

Fig. \ref{f12} shows the IPR $L(\omega)$ of the lowest energy
excitations for $U/t=12$ and various values of the disorder. For
weak $\Delta/t=1.5$ (Fig. \ref{f12} A), the majority of excitations
are delocalized, as expected, with a localized `tail'  at
high $\omega$. Increasing disorder to $\Delta/t=3$ and 4.5 leads to the
occurrence of further sparse, localized excitations at very low $\omega$,
superimposed over a background of delocalized excitations (Fig. \ref{f12}
B,C); and the density of localized excitations rises on
further increasing disorder (Fig. \ref{f12} D).

The above behaviour  is clarified by considering the corresponding IPR 
probability
distributions, $P(L)$ (Fig. \ref{f13}). 
This shows a peak at low $L$, corresponding to the `delocalized'
backbone, with a long tail to high $L$, reflecting the presence of
localized excitations. For low $\Delta/t=1.5$, the peak is sharp and
centred well within our approximate criterion for delocalized
excitations. As $\Delta/t$ is increased, the 
distribution broadens and the most probable value of
$L$ shifts to higher values. 

To understand the above behaviour, we 
site-resolve excitations of given IPR $L$, in analogy to the site
distribution of quasiparticle states, $H(\epsilon;E)$  (Eq. (\ref{heps})), via the quantity
\begin{equation}
H_{\rm RPA}(\epsilon;L)={g(\epsilon)N_{\hbox{\tiny RPA}}(\epsilon;L)\over
N_{\hbox{\tiny RPA}}(L)}
\label{overlap}
\end{equation}
(such that $\int d\epsilon H_{\rm RPA}(\epsilon;L)=1$). Here,
$N_{\hbox{\tiny
RPA}}(L)=N^{-1}\sum_{\omega_p}\delta(L-L(\omega_p))$ is the density of
excitations of fixed IPR $L$, and $N_{\hbox{\tiny RPA}}(\epsilon;L)$ is
the partial density of such excitations on sites of given
site energy $\epsilon$, given by

\begin{equation}
N_{\hbox{\tiny RPA}}(\epsilon;L) =
N_\epsilon^{-1}\sum_{i:\epsilon_i=\epsilon}\limits
\sum_{\omega_p}\limits |V_{i\gamma}(\omega_p)|^2\delta(L-L(\omega_p)).
\end{equation}
$H_{\rm RPA}(\epsilon;L)$ then gives the overlap of excitations of given
IPR $L$ on sites of different site energies.
For $(\Delta/t,U/t)=(3,12)$, Fig. \ref{f14} shows the $\epsilon$-dependence of
$H_{\rm RPA}$ averaged over all $L<0.1$ and $L>0.3$, to illustrate the
site-differential character of extended and localized excitations
respectively. Localized excitations are seen to be dominated by sites close to the
local moment boundaries $|\epsilon|\sim U/2$, 
while sites in the range $|\epsilon|\lesssim
U/2$, which carry large moments, make the dominant
contribution to delocalized excitations. As expected, sites outside
the local moment regime, which possess very small moments, do not
contribute significantly to any of the low energy excitations;
rather, they participate  in Stoner-like excitations of significantly
higher energy. 

The origin of the localized and delocalized low energy transverse spin 
excitations is then readily rationalized. Sites with site energies in the range
$|\epsilon|\lesssim U/2$, which possess large local moments, contribute
most significantly to single particle states within the Hubbard
sub-bands. Since these states are the most delocalized,  
collective excitations involving them will likewise be delocalized. 
At UHF (`pure Stoner') level, transverse spin excitations between 
single-particle states within the Hubbard bands incur an energy cost of
order $U$; but in the RPA spectrum, via inclusion of particle-hole
correlations, this energy is renormalized down to a spin wave-like scale
on the order of an exchange coupling. This is the origin of the
delocalized low-energy transverse spin excitations. By contrast, sites
with $|\epsilon|\sim U/2$ close to the local moment boundaries
participate dominantly in single-particle states close to the Fermi
level, or an energy $\sim U$ above or below it. For $U/t=12$ such states
are strongly localized, hence so too are collective excitations
involving them. Again, such excitations are renormalized to a low-energy
scale via the particle-hole interactions inherent in the RPA. Both
localized and delocalized transverse spin excitations thus arise at low
energies, reflecting the underlying localization characteristics of single
particle states over a very wide energy range in the associated
single-particle spectrum. Note further that these arguments do not imply
coexistence of localized and extended transverse spin excitations at the
same energy in the spin wave-like band; and the apparent occurrence of
such in eg Fig. 7D is largely 
a consequence of sampling different disorder realizations.
While the present, qualitative considerations do not of course permit
identification of mobility edges in the spin wave-like band, it is at
least plausible from Fig. 7 that such exist.

From the above, the system can thus  be regarded as composed of two
essential components: an AF large-moment component which gives rise
to the backbone of delocalized excitations, and a set of sites
carrying smaller moments which give rise to localized excitations. As
disorder is increased, the fraction of sites in the former category
decreases, leading to a shift in the peak in $P(L)$ (Fig. \ref{f13}) to
higher $L$ and a concomitant broadening of the distribution. The probability
of sites with $|\epsilon|\sim U/2$ correspondingly increases,
and hence localized excitations ultimately become the dominant theme.

The above discussion is confined to a region of the ($\Delta/t,U/t$) phase
diagram in which the mean field ground state is  an AF insulator. What of the
metallic AF and SG phases? Fixing $\Delta/t=3$ and decreasing $U/t$
progressively from $U/t=12$ down to $U/t=6$ leads (at $U/t\simeq 7.2$)
 to a transition from an AF
insulator to an AF metal. However, the IPR profile of the low-energy transverse spin
excitations shows essentially no qualitative change from that for $U/t=12$: a backbone
of delocalized excitations remains present, with localized excitations
superimposed at low and higher $\omega/t$. In the metallic region, however, 
the localized excitations now arise principally from transfer of a spin 
between a {\em delocalized} UHF state near the Fermi level to
a {\em localized} UHF state $\sim U$ above or below $E_F$. The localized
state overlaps only a few sites, while the delocalized state has weight on a
finite fraction of sites. The excitation thus appears localized in the IPR
profile since the on-site spin flip contribution 
(which is of course the component of the excitation probed, see Eq.
(2.14)) can occur only from sites overlapped by the localized state. 

On decreasing $U/t$ further to $U/t\simeq 2.5$, the mean field ground
state becomes a SG metal. Here, the RPA and pure Stoner excitation spectra
appear virtually identical (see e.g. Fig. \ref{f9}).
It is indeed found that the majority of low energy RPA excitations are
little more than weakly renormalized Stoner excitations, as expected
from the low value of $\bar{\Omega}$ in Table I. However, at very low
energies ($\omega\lesssim 0.3t$), this is 
not the case. As in the AF metal regime,
low energy localized modes appear in the RPA spectrum, while the pure
Stoner 
spectrum at such low energies is composed of entirely delocalized
modes (since the system is metallic). These localized excitations are found to
have dominant weight on those rare sites in the spin glass that possess
substantial local moments, flips of which would incur an
energy cost of $\sim U$ within UHF, but occur at significantly lower
energies in the RPA spectrum.

\section{Discussion}

In recent papers \cite{prbhub,prl}, we have described an approximate
mapping of the low-frequency transverse spin excitations of the
non-disordered Hubbard model at finite $U/t$, on to those of an
effective underlying Heisenberg model. 
The $U/t$-dependent effective exchange couplings are not 
restricted to purely nearest neighbour interactions; they follow
solely from a knowledge of the RPA excitations about the broken-symmetry
UHF state at zero temperature.
The mapping
becomes exact as $U/t\rightarrow\infty$ and, conjoined with an Onsager
reaction field approach, enables extraction of thermodynamic
properties in the thermal paramagnetic phase. The resultant N\'eel
temperatures, spin correlation functions and magnetic susceptibilities
were  \cite{prbhub,prl} found to be in good agreement with Quantum
Monte Carlo calculations over a wide $U/t$-range from strong to weak
coupling. We now assess the extent to which the present findings
support application of such a mapping to the Anderson-Hubbard model.

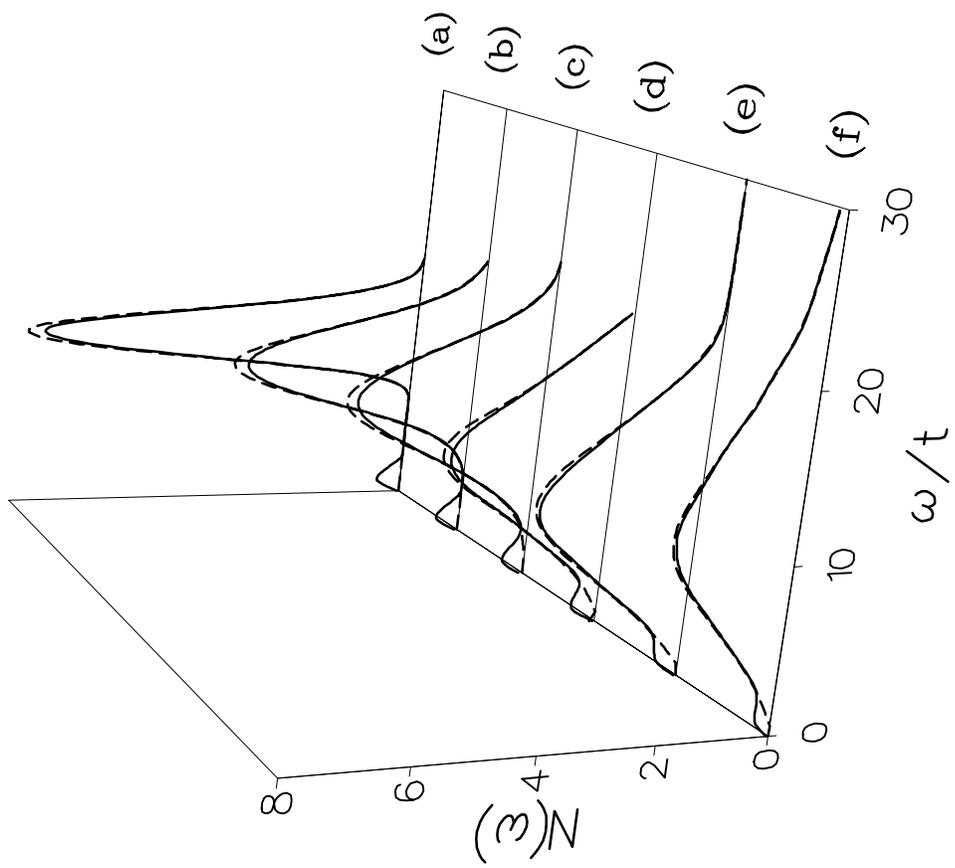
\begin{figure}
\caption{
As Fig. \protect\ref{f9} but with fixed interaction strength $U/t=12$ and
$\Delta/t=1$ (a), 2 (b), 3 (c), 4.5 (d), 6 (e) and 7.5 (f). 
Note the persistence of a spin
wave-like band for all disorder strengths.
}
\label{f10}
\end{figure}

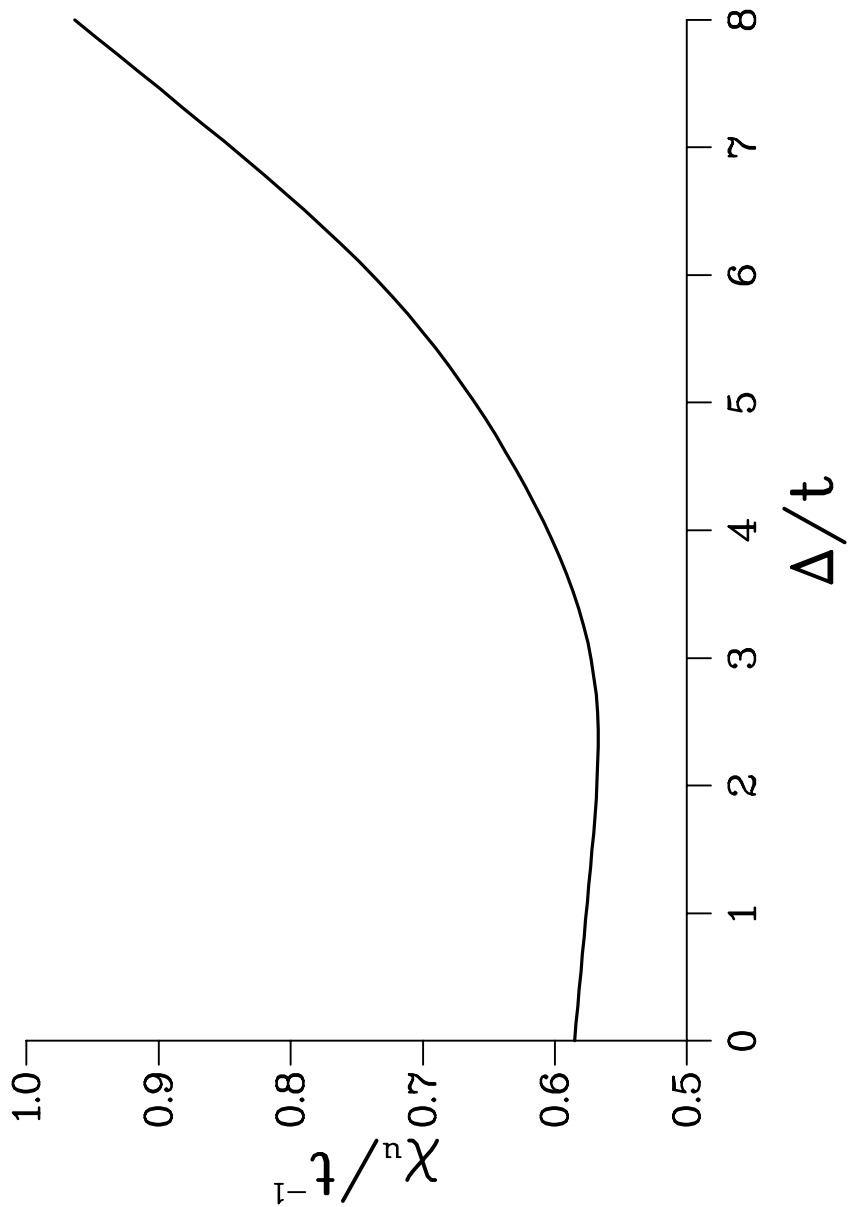
\begin{figure}
\caption{
Uniform RPA static transverse susceptibility $\chi_u$ for fixed $U/t=12$
as a function of disorder $\Delta/t$.
}
\label{fchi}
\end{figure}

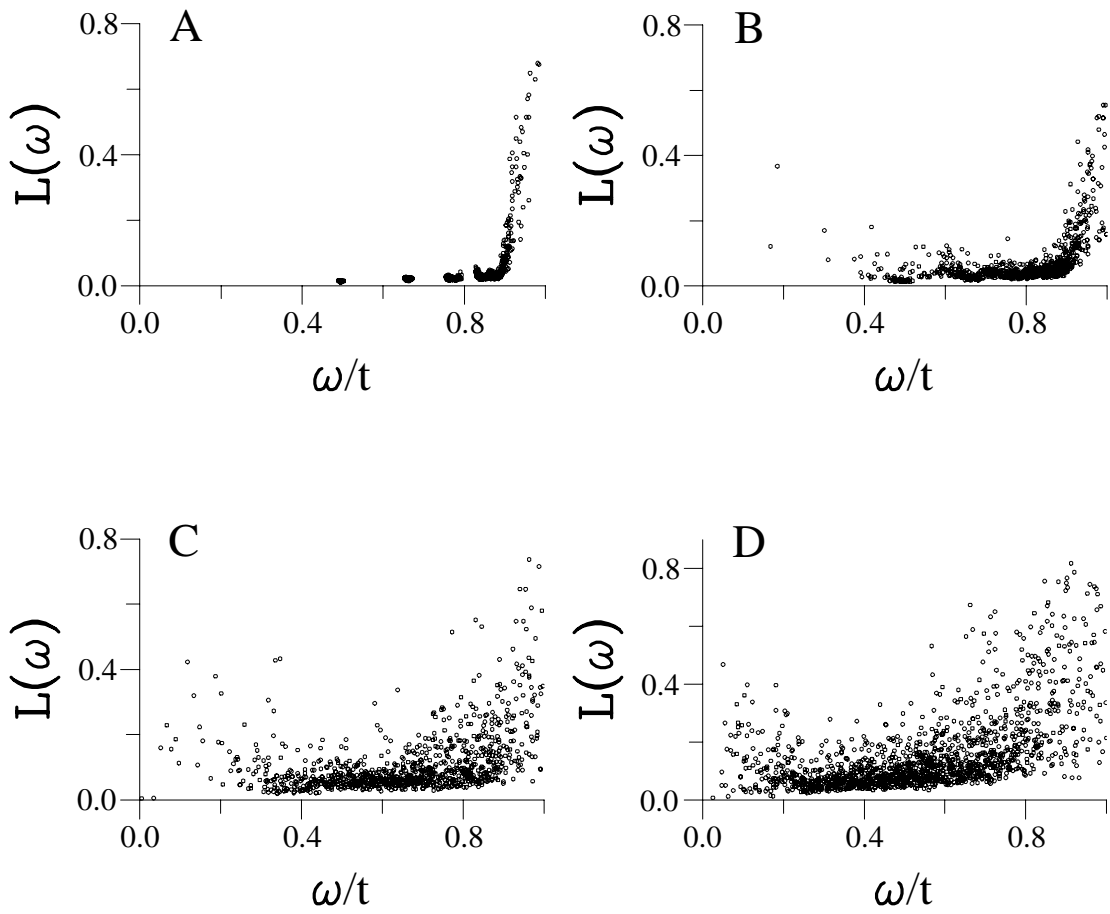
\begin{figure}
\caption{
Scatter diagram showing IPR $L$ for low-frequency RPA excitations vs $\omega/t$
for fixed interaction strength $U/t=12.0$ and varying disorder, $\Delta/t=1.5$
(A), 3.0 (B), 4.5 (C), 6.0 (D).
}
\label{f12}
\end{figure}

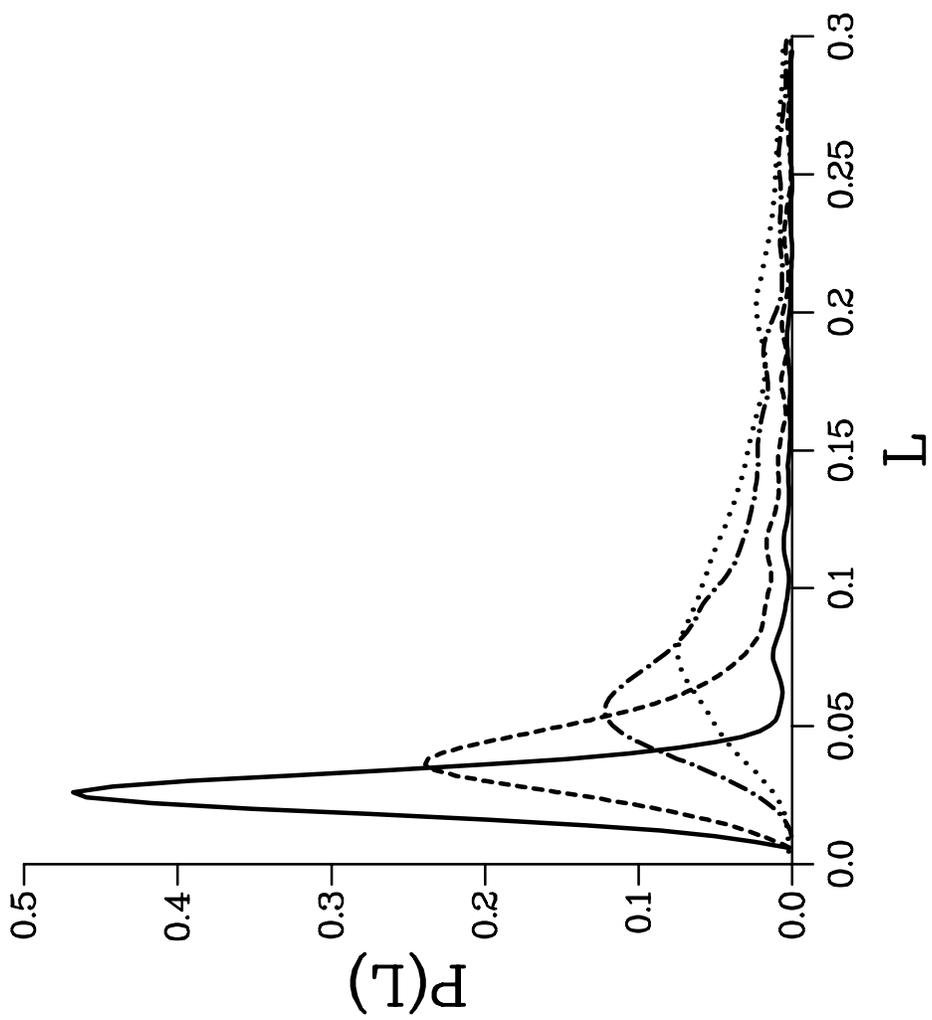
\begin{figure}
\caption{
Probability distribution of IPRs $P(L)$ for $U/t=12$ and $\Delta/t=1.5$
(solid), 3.0 (dashed), 4.5 (dot-dashed) and 6.0 (dotted).
}
\label{f13}
\end{figure}

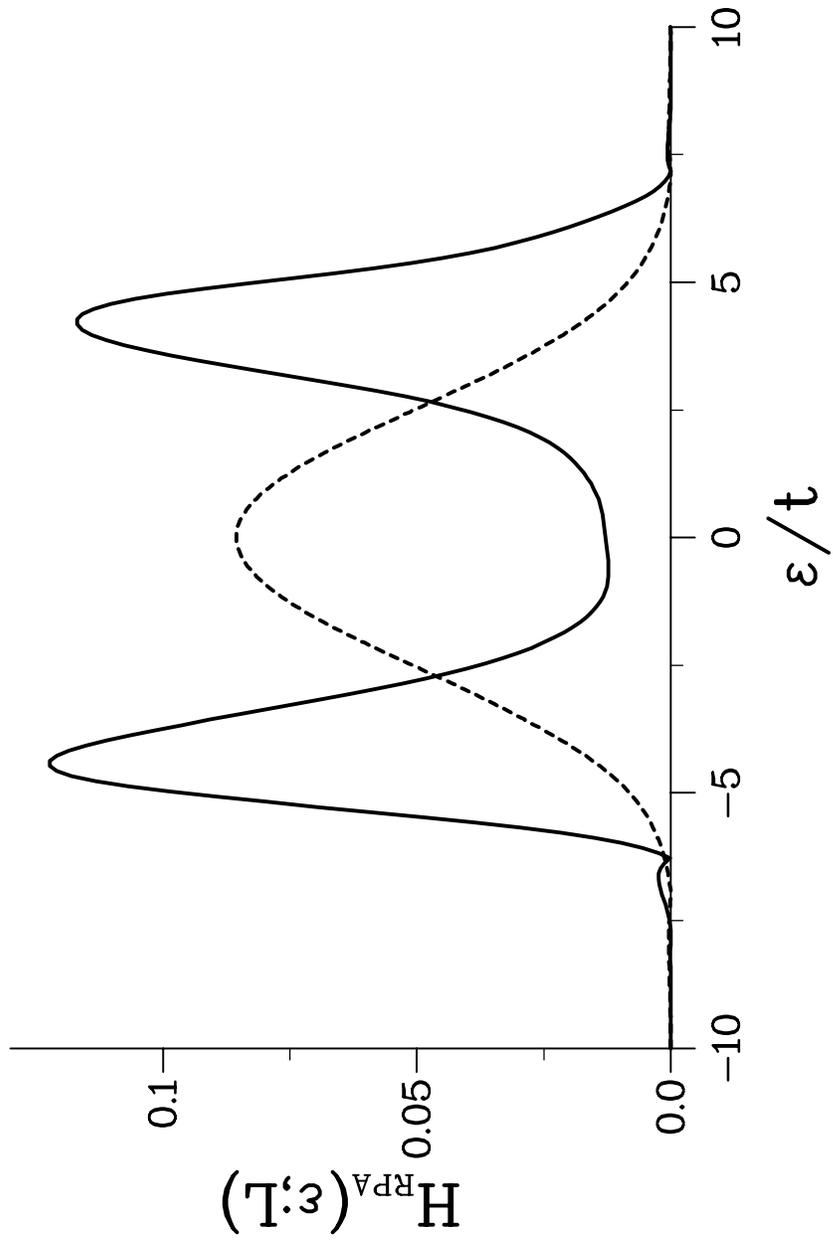
\begin{figure}
\caption{Probability distribution $H_{\rm RPA}(\epsilon;L)$ (Eq. (\protect\ref{overlap}))
of the lowest $N$  RPA excitations with IPR $L$ over sites of energy $\epsilon$, averaged over
excitations with $L>0.3$ (solid line) and $L<0.3$ (dotted line) for
$U/t=12$ and $\Delta/t=3$.                                    
}
\label{f14}
\end{figure}

As discussed in Refs  \onlinecite{prbhub,prl}, the mapping is accurate
provided there is a persistent separation between low energy spin
wave-like excitations and higher energy Stoner-like excitations;
i.e. provided in practice a discernible spin wave-like band,
comprising excitations
with an appreciable degree of on-site spin flip character, is present in the full RPA
transverse spin spectrum. As shown in \S III,
this indeed holds for the pure Hubbard model for all  $U/t\gtrsim 2-3$.
For the Anderson-Hubbard model, Fig. \ref{f10} and Table I show that this also
holds for all disorder strengths studied, at relatively large
$U/t=12$. Thus, the mapping should certainly be accurate well within the
disordered AF phase of the model. On fixing disorder and reducing
interaction strength towards the AF--SG boundary, the spin wave character
of the low-frequency band decreases significantly and the band itself
is ultimately absorbed by the Stoner-like portion, such that for $\Delta/t=3$
the mapping would not be quantitatively accurate for
$U/t\simeq 4$ and below. Not unexpectedly, therefore, the mapping
would not be applicable to the low-$U/t$ SG phase; but we do
anticipate it to be accurate throughout the major part of the
disordered AF phase in the $(\Delta/t,U/t)$ plane, wherein the effect
of disorder on, for example, the N\'eel temperature 
can be investigated. This is the subject of a forthcoming paper \cite{pap}.

\acknowledgments

We are grateful to the EPSRC (Condensed Matter Physics) for financial
support.

\mediumtext
\begin{table}
\caption{
Spin wave character of low-$\omega$ RPA excitations $\Omega$ for the
non-disordered and disordered Hubbard models. For $\Delta/t=0$,
$\Omega=\Omega(\omega_{\rm max})$; $\bar{\Omega}$ is the average of
$\Omega(\omega)$ over the lowest $N$ excitations.
}
\medskip
\begin{tabular}{cccccc}
\multicolumn{2}{c}{$\Delta/t=0$}&\multicolumn{2}{c}{$\Delta/t=3.0$} & 
\multicolumn{2}{c}{$U/t=12$} \\
$U/t$ &  $\Omega$ & $U/t$ & $\bar{\Omega}$ & $\Delta/t$ & $\bar{\Omega}$ \\
\tableline
1.0 & 0.02  & 3.0 & 0.05 & 0.0 & 1.00 \\ 
2.0 & 0.39 & 4.5 & 0.20 & 1.5 & 1.00 \\
3.0 & 0.71  & 6.0 & 0.33 & 3.0 & 0.98 \\
4.0 & 0.86 & 9.0 & 0.75 & 4.5 & 0.89 \\
$\geq$ 5.0 & 1.00 & 12.0 & 0.98 & 6.0 & 0.88 
\end{tabular}
\label{table1}
\end{table}

\end{document}